%% file: main.tex
\PassOptionsToPackage{table}{xcolor}
\documentclass[10pt,twocolumn,letterpaper]{article}

\usepackage{cvpr}
\usepackage{placeins}
\usepackage{subcaption}
\usepackage[percent]{overpic}
\usepackage{siunitx}
\usepackage{amsfonts}
\usepackage{amsmath}
\usepackage{graphicx}
\usepackage{xcolor}
\usepackage{booktabs}
\usepackage{colortbl}

\usepackage{dblfloatfix}

\newcommand{\shortcite}{
	\cite
}

\newcommand{\cfbox}[2]{%
	\colorlet{currentcolor}{.}%
	{\color{#1}%
		\fbox{\color{currentcolor}#2}}%
}

\usepackage[breaklinks=true,bookmarks=false]{hyperref}

\cvprfinalcopy 

\def\cvprPaperID{****} 

\graphicspath{{.}{./figures/}{"G:/Team Drives/Stanford Computational Imaging/Projects/HDR_ONN/figures/"}{"G:/Team Drives/Stanford Computational Imaging/Projects/HDR_ONN/"}}

\begin{document}
	
	\title{Deep Optics for Single-shot High-dynamic-range Imaging}
	
	\author{Christopher A. Metzler\hspace{1 cm} Hayato Ikoma \hspace{1 cm}Yifan Peng \hspace{1 cm}Gordon Wetzstein \\
		Stanford University\\
		{\tt\small \{cmetzler, hikoma, evanpeng, gordon.wetzstein\}@stanford.edu}
	}

%
%
%
	
	\maketitle

%

%
\begin{abstract}
High-dynamic-range (HDR) imaging is crucial for many computer graphics and vision applications. Yet, acquiring HDR images with a single shot remains a challenging problem. Whereas modern deep learning approaches are successful at hallucinating plausible HDR content from a single low-dynamic-range (LDR) image, saturated scene details often cannot be faithfully recovered. Inspired by recent deep optical imaging approaches, we interpret this problem as jointly training an optical encoder and electronic decoder where the encoder is parameterized by the point spread function (PSF) of the lens, the bottleneck is the sensor with a limited dynamic range, and the decoder is a convolutional neural network (CNN). The lens surface is then jointly optimized with the CNN in a training phase; we fabricate this optimized optical element and attach it as a hardware add-on to a conventional camera during inference. In extensive simulations and with a physical prototype, we demonstrate that this end-to-end deep optical imaging approach to single-shot HDR imaging outperforms both purely CNN-based approaches and other PSF engineering approaches.
\end{abstract}

\begin{figure*}
	\centering
	\vspace{1 cm}
		\begin{subfigure}[b]{.21\textwidth}
			\centering
			\captionsetup{justification=centering}
			\begin{overpic}[width=\textwidth]{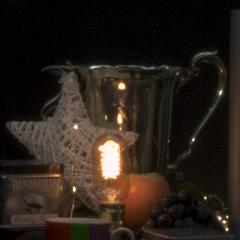}
				\put (25,105) {{\bf LDR Image}}
			\end{overpic}
			\caption*{0 EV}
		\end{subfigure}
		~
		\begin{subfigure}[b]{.21\textwidth}
			\centering
			\captionsetup{justification=centering}
			\begin{overpic}[width=\textwidth]{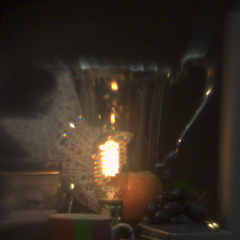}
				\put (5,105) {{\bf E2E Measurement}}
			\end{overpic}
			\caption*{0 EV}
		\end{subfigure}
		~
		\begin{subfigure}[b]{.21\textwidth}
			\centering
			\captionsetup{justification=centering}
			\begin{overpic}[width=\textwidth]{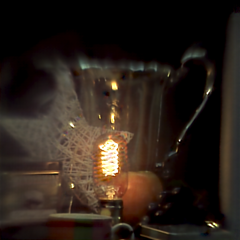}
				\put (0,105) {{\bf E2E Reconstruction}}
			\end{overpic}
			\caption*{0 EV}
		\end{subfigure}
		~
		\begin{subfigure}[b]{.0755\textwidth}
		\centering
		\captionsetup{justification=centering}		
		\includegraphics[width=0.95\textwidth]{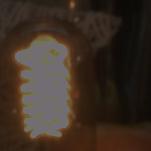}
		\includegraphics[width=0.95\textwidth]{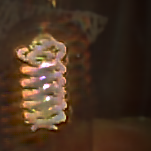}
		\begin{overpic}[width=0.95\textwidth]{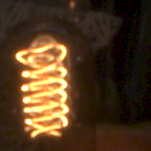}
		\end{overpic}
		\caption*{-2.3 EV}
	\end{subfigure}
	~	
	\begin{subfigure}[b]{.21\textwidth}
		\centering
		\captionsetup{justification=centering}
		\begin{overpic}[width=\textwidth]{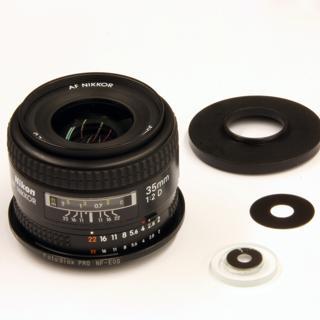}
			\put (20,105) {{\bf Optical Setup}}
		\end{overpic}
		\caption*{}
	\end{subfigure}
  \caption{Conventional camera sensors are limited in their ability to capture high-dynamic-range (HDR) scenes. Details in brighter parts of the image, such as the light bulb, are saturated in a low-dynamic-range (LDR) photograph (left). We propose an end-to-end (E2E) approach to jointly optimizing a diffractive optical element and a convolutional neural network (CNN) to enable single-shot HDR imaging. After training optics and algorithm jointly, the lens is fabricated and attached to a conventional camera lens as an add-on (right). During inference, the proposed deep optical imaging system records a single sensor image (center left) that contains optically encoded HDR information, which helps the CNN recover an HDR image (center right). As compared to the conventional LDR image (insets, top), the HDR image computed by our system (insets, center) extends the dynamic range of the sensor significantly, and more closely resembles the reference HDR photograph of this scene (insets, bottom).
	}
  \label{fig:teaser}
\end{figure*}


%
\maketitle

\input{sections/definitions}

\section{Introduction}
\label{sec:intro}
\input{sections/introduction}

\section{Related Work}
\label{sec:related}
\input{sections/related}

\section{End-to-end HDR Imaging}
\label{sec:methods}

\input{sections/methods}

\section{Analysis and Evaluation}
\label{sec:analysis}
\input{sections/analysis}

\section{Fabrication and Implementation}
\label{sec:prototype}
\input{sections/prototype}

\section{Experimental Results}
\label{sec:results}
\input{sections/results}

\section{Discussion}
\label{sec:discussion}
\input{sections/discussion}

\section{Conclusion}
\label{sec:conclusion}
\input{sections/conclusion}


{\small
	\bibliographystyle{ieee_fullname}
	\bibliography{MyBib}
}

\end{document}

%% file: sections/definitions.tex
\newcommand{\myparagraph}{\noindent\textbf}

\newcommand{\cc}{f}

\newcommand{\noise}{{\boldsymbol{\eta}}}

\newcommand{\psf}{{\mathbf{h}}}

\newcommand{\meas}{{\mathbf{y}}}

\newcommand{\img}{{\mathbf{x}}}

\newcommand{\rec}{G}

\newcommand{\doe}{\boldsymbol{\phi}}

\newcommand{\doephase}{\mathbf{t}_{\phi}}

\newcommand{\lensphase}{\mathbf{t}_{l}}

\newcommand{\focallength}{g}

\newcommand{\note}[1]{\textcolor{red}{#1}}

%% file: sections/introduction.tex
High dynamic range (HDR) imaging is one of the most widely used computational photography techniques with a plethora of applications, for example in image-based lighting~\cite{Debevec:IBL}, HDR display~\cite{Seetzen:2004}, and image processing~\cite{reinhard2010high,Banterle:2011}. However, the dynamic range of a camera sensor is fundamentally limited by the full well capacity of its pixels. When the number of generated photoelectrons exceed the full well capacity, which is typically the case when imaging scenes with a high contrast, intensity information is irreversibly lost due to saturation. Ever shrinking pixel sizes, for example in mobile devices, exacerbate this problem because the full well capacity is proportional to the pixel size. 

Several different strategies have been developed to overcome the limited dynamic range of available sensors. One class of techniques captures multiple low-dynamic-range (LDR) sensor images with fixed~\cite{HDRplus} or varying~\cite{Picard95onbeing,Debevec:1997,mertens2009exposure} exposure settings. Unfortunately, motion can be problematic when capturing dynamic scenes with this approach. Another class of techniques uses multiple optically aligned sensors~\cite{McGuire:2007,Tocci:2011} to capture these exposures simultaneously, but calibration, cost, and device form factor can be challenging with such special-purpose cameras. Single-shot approaches are an attractive solution, but traditionally required custom exposure patterns to be multiplexed on the sensor~\cite{Nayar:2000,Hajisharif2015,Serrano:2016}. Most recently, single-shot HDR imaging approaches were proposed that hallucinate an HDR image from a single saturated LDR image (HDR-CNN, e.g.~\cite{HDRCNN}). While successful in many cases, saturated scenes details often cannot be faithfully recovered via hallucination. 

In this work, rather than hallucinating missing pixel values, we aim to preserve information about the saturated pixel values by encoding information about the brightest pixel values into nearby pixels via an optical filter with an optimized point spread function (PSF). Unlike previous attempts to encode HDR pixel information with an optical filter~\cite{Rouf:2011}, we turn to machine learning to automatically design both the optical element and the reconstruction algorithm end-to-end, so as to maximize the information passed from the HDR scene to the low-dynamic-range (LDR) measurements. In essence, we construct an autoencoder where the encoding is performed optically and the decoding is performed computationally. Both encoder and decoder are trained in an end-to-end fashion, with the optimized optical element being fabricated and remaining fixed during inference.
 
In optimizing the encoder and decoder, our system must solve three challenging inverse problems at once. (1) Mapping a PSF to a manufacturable optical filter is implicitly a phase retrieval problem. (2) Using optically encoded information to fill in saturated regions is an inpainting problem. (3) Removing said optically encoded information from non-saturated regions is a deconvolution problem. Our work is the first to explore and successfully address this unique and challenging combination of problems. 

Using extensive simulations, we demonstrate that deep optics generally achieves better results than alternative single-shot HDR imaging approaches. This is intuitive, because compared with HDR-CNN approaches, our optimized PSF has more degrees of freedom to encode scene information in the sensor image, and compared with other optical encoding techniques, ours uses an optical element that is jointly optimized with the reconstruction algorithm, rather than heuristically chosen. We demonstrate the proposed camera system with a proof-of-concept prototype by fabricating a diffractive optical element that can simply be attached as a hardware add-on to a conventional camera lens. 

Specifically, we make the following contributions
\begin{itemize}
	\item We introduce an optical encoder and CNN-based decoder pipeline for single-shot HDR imaging. 
	\item We present a new single-shot ``multiplexing'' approach to HDR imaging; the learned, grating-like diffractive optical element (DOE) creates shifted and scaled copies of the image which are used to reconstruct the brightest regions of the scene.
	\item We analyze the proposed system and demonstrate that it outperforms existing single-shot HDR methods.
	\item We fabricate the optimized diffractive optical element and validate the proposed system experimentally.
\end{itemize}

\paragraph{\bf Overview of Limitations}

The proposed approach to single-shot HDR imaging via deep optics is successful in many scenarios. However, it does make computational processing an integral part of the image formation, which may increase the computational burden compared to conventional LDR imaging. Similarly to other single-shot approaches, ours may fail to robustly estimate a high contrast scene for extremely large saturated regions in the measurements.



%% file: sections/related.tex
\paragraph{\bf HDR Imaging}

High dynamic range imaging aims at overcoming the limited dynamic range of conventional image sensors using computational photography techniques. Many approaches rely on capturing several LDR images with different exposures and fusing them into a single HDR image~\cite{Picard95onbeing,Debevec:1997,mertens2009exposure,HDRplus,Hasinoff:2010}. Although motion between the LDR images can be a problem, many proposals have been introduced to deal with this problem~\cite{Khan:2006,Gallo:HDRmotion,Granados:2013,Hu:2013,LearningHDR}, allowing even HDR video to be recorded from temporally varying exposures~\cite{Kang:2003,Sen:2012:RPH,HDRVideo}. 

Although many of these multi-shot HDR imaging approaches are successful in some scenarios, they can fail for fast motion and they can also be computationally expensive. To mitigate these limitations, multiple sensors can be optically combined to capture these exposures simultaneously~\cite{DBLP:journals/ijcv/AggarwalA04,McGuire:2007,Tocci:2011}. This approach, however, is costly and bulky and the system calibration can be challenging. 

Motivated by these shortcomings, several approaches to single-shot HDR imaging have been proposed. Such reverse tone mapping approaches aim at solving an ill-posed problem~\cite{Banterle:2006,Meylan:85838,Rempel:2007}. Using computational photography approaches, this problem can be made less ill-posed, for example using spatially varying pixel exposures via neutral density filter arrays or spatially varying ISO settings~\cite{Nayar:2000,5540166,Hajisharif2015,Serrano:2016}, using an optically coded point spread function (PSF)~\cite{Rouf:2011}, or using a special modulo camera~\cite{Zhao:2015}. Most recently, convolutional neural networks (CNNs) have been employed to hallucinate realistic HDR images from a single LDR image~\cite{HDRCNN,Endo:2017,lee2018deep}. 

Our work also uses a CNN to recover an HDR image from a single LDR image, but rather than hallucinating it, we use an optimized PSF to encode as much of the HDR image content as possible in the sensor image. While Rouf et al.~\shortcite{Rouf:2011} also used an optical filter to aim for the same goal, theirs was heuristically chosen and is limited in its ability to recover high-quality HDR images. We train a CNN end-to-end with an optimized optical filter that achieves far superior image quality. We fabricate this optimized lens using grayscale lithography and demonstrate its ability to capture single-shot HDR images with a prototype camera system. 

\begin{figure*}[t]
	\centering
	\includegraphics[width=.95\linewidth]{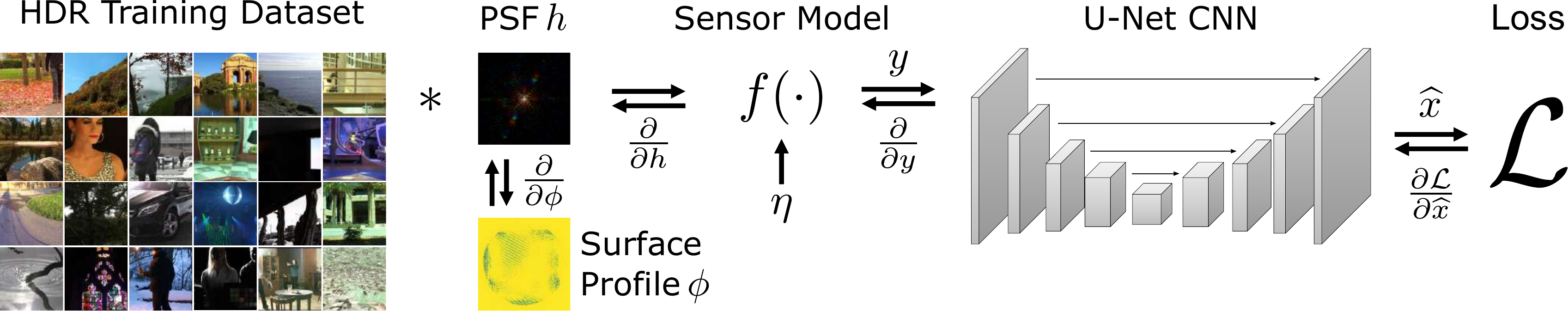}
	\caption{Illustration of the proposed end-to-end optimization framework. HDR images of a training set are convolved with the PSF created by a lens surface profile $\doe$. These simulated measurements are clipped by a function $\cc (\cdot)$ to emulate sensor saturation and noise $\noise$ is added. The resulting RGB image $\meas$ is processed by a convolutional neural network (CNN) and its output compared with the ground truth HDR image using the loss function $\mathcal{L}$ described in the text. In the learning stage, this loss is back-propagated into the CNN weights and bias values and also into the height values $\doe$ of the lens. During inference, a captured LDR image blurred by the optical PSF is fed directly into the pre-trained CNN to compute the reconstructed HDR image.}
	\label{fig:SystemDiagram}
\end{figure*}

\paragraph{\bf Computational Optics}

There is a long history of co-designing optics and image processing. In computational photography, research in this topic has focused on various applications such as extended-depth-of-field imaging~\cite{Dowski:95,Cossairt:2010:DCP,Cossairt:2010}, motion or defocus deblurring~\cite{Raskar:2006,Zhou:2009,Zhou:2009b}, depth estimation~\cite{Levin:2007,Levin:2009}, multispectral imaging~\cite{Wagadarikar:08,Choi:2017,jeon2019compact}, light field imaging~\cite{Ng:2005,Veeraraghavan:2007,Marwah:2013}, achromatic imaging~\cite{Peng:2016}, gigapixel imaging~\cite{Cossairt:2011,Brady:2012}, and lensless imaging~\cite{Antipa:2016,Asif:2017}. In computational microscopy, similar concepts are known as point spread function (PSF) engineering and have been used for optimizing the capabilities of single-molecule localization microscopy~\cite{Pavani:2009,Shechtman:2014}. In all of these examples, some optimality criterion is defined for the PSF, which is then optimized to work well for a particular choice of algorithm. This can be interpreted as co-design, whereas our approach builds on the emerging concept of end-to-end design, where optics and image processing are optimized jointly in an end-to-end fashion. 


\paragraph{\bf Deep Optics}

The idea of end-to-end optimization of optics and image processing has recently gained a lot of attention. This concept has been demonstrated to provide significant benefits for applications in color imaging and demosaicing \cite{chakrabarti2016learning}, extended depth of field and superresolution imaging~\cite{VincentEndtoEnd}, monocular depth imaging~\cite{he2018learning,Haim:2018,Wu:2019,chang2019deep}, image classification~\cite{chang2018hybrid}, time-of-flight imaging~\cite{Marco:2017:DOR,Su_2018_CVPR}, computational microscopy~\cite{horstmeyer2017convolutional, Hershko:19,nehme2019dense,kellman2019data}, and focusing light through scattering media \cite{turpin2018light}. 
To the best of our knowledge, this work is the first to explore deep optics for single-shot HDR imaging. We perform a detailed evaluation of this idea for HDR imaging and demonstrate practical benefits with a custom prototype camera.

%% file: sections/methods.tex

A camera maps a scene $\img$ to a two-dimensional sensor image $\meas$ as
\begin{equation}
	\meas = \cc \left( \psf * \img + \noise \right),
\end{equation}
where $\img \in \mathbb{R}_+^{n_x \times n_y}$ is a discrete image with $n_x \times n_y$ pixels, each containing values that are proportional to the irradiance incident on the sensor. The irradiance is scaled such that the non-saturated values map to the range $\img_i \in [0,1]$. Furthermore, $\noise$ models signal-independent read noise, $\psf$ is the optical point spread function (PSF) created by the camera lens, $*$ denotes the 2-D convolution operator, and $\cc(\cdot)$ is the camera's response function. This image formation model assumes that the PSF is shift-invariant, but the model could be generalized to describe PSFs that vary laterally or with depth. 

We assume that the camera has a linear camera response function, which is typically the case when working with raw sensor data:
\begin{equation}
	\cc \left( \img_i \right) =\begin{cases}	
	0, 			& \text{ if } \img_i<0, \\
	\img_i, & \text{ if } 0 \leq \img_i \leq 1,\\
	1, 			& \text{ if }\img_i> 1.
	\end{cases}
\end{equation}
Nonlinear camera response functions can be calibrated and inverted so as to mimic a linear response function~\cite{Mitsunaga:99}. We ignore the effects of quantization.

Our goal in this work is to jointly optimize the PSF $\psf$ and a reconstruction algorithm $\rec : \mathbf{y} \mapsto \hat{\img}$ so as to recover $\img$ from $\meas$ when $\|\img\|_\infty \gg 1$. To this end, we turn to differentiable optical systems and algorithms, which we describe in the following.

\subsection{Modeling the Optical Point Spread Function}

As shown in Figure~\ref{fig:teaser}, our optical system is a conventional single lens reflex camera lens with a custom diffractive optical element (DOE) add-on. Similar to a photographic filter, the DOE is mounted directly on the lens. To model the light transport from a scene, through these optical elements, to the sensor, we build on a differentiable Fourier optics model~\cite{goodman2005introduction}, an approach closely related to recent work on end-to-end camera designs~\cite{VincentEndtoEnd,chang2018hybrid,Wu:2019,chang2019deep}. Specifically, our aim is to find the microscopic surface profile $\doe$ of the DOE that creates a PSF $\psf$ which is optimally suited for the HDR image reconstruction algorithm. 


Assuming that the scene is at optical infinity, the complex-valued wave field of a point located on the optical axis becomes a plane wave immediately before the DOE, i.e. $\mathbf{u}_{in} = \exp \left( i k z \right)$, where $k=\frac{2 \pi}{\lambda}$ is the wave number and $\lambda$ is the wavelength. The phase of this wave is affected by the DOE in a spatially varying manner by a complex-valued phase delay $\doephase$, which is directly calculated from the surface profile $\doe$ as 
\begin{equation}
	\doephase (u,v,\lambda) = \mathbf{A}_{\phi} \left( u,v \right) \cdot \exp \left( i k \left(n( \lambda)-1 \right) \doe (u,v) \right).
\end{equation}
Here, $u,v$ are the lateral coordinates on the DOE surface, $n( \lambda)$ is the wavelength-dependent refractive index of the material that the DOE is made of and $A_{\doe} \left( u,v \right)$ is a binary circular mask with diameter $D_\phi$ that models the aperture of the DOE as 
\begin{equation}
\mathbf{A}_{\doe}(u,v)=\begin{cases}
	1, & \text{if} $\,\,$ u^2+v^2 \leq \left(\frac{D_\phi}{2} \right)^2,\\
	0, & \text{otherwise.}
	\end{cases} 
\end{equation}
The wave field continues to propagate by some distance $d_{\phi}$ to the camera lens with focal length $\focallength$. This lens induces the following phase delay:
\begin{align}
	\lensphase \left( u',v' \right) = \mathbf{A}_{l} \left( u',v' \right) \cdot \exp \left( -i \frac{k}{2 \focallength} \left( u'^2+v'^2 \right) \right).
	\label{eqn:ThinLensPhaseDelay}
\end{align}
Although the physical compound lens contains many optical elements that correct distortions and chromatic aberrations, using a simplified thin lens model (Eq.~\ref{eqn:ThinLensPhaseDelay}) adequately describes the mathematical behavior of the lens.

Finally, the wave field propagates by a distance $d_s$ to the sensor, where its intensity $\psf = \left| \mathbf{u}_{sensor} \right|^2$ is recorded. Putting all of this together results in  
\begin{equation}
	\psf_{\doe} \left( x,y \right) = \left| P_{d_s} \left\{ \lensphase \cdot P_{d_{\phi}} \left\{ \doephase \cdot exp \left( i k z \right) \right\} \right\} \right|^2,
\end{equation}
where $P_d \left\{ u \right\}$ models free-space propagation of a wave field $u$ by a distance $d$.

\begin{figure}[t!]
\centering
	\includegraphics[width=\linewidth]{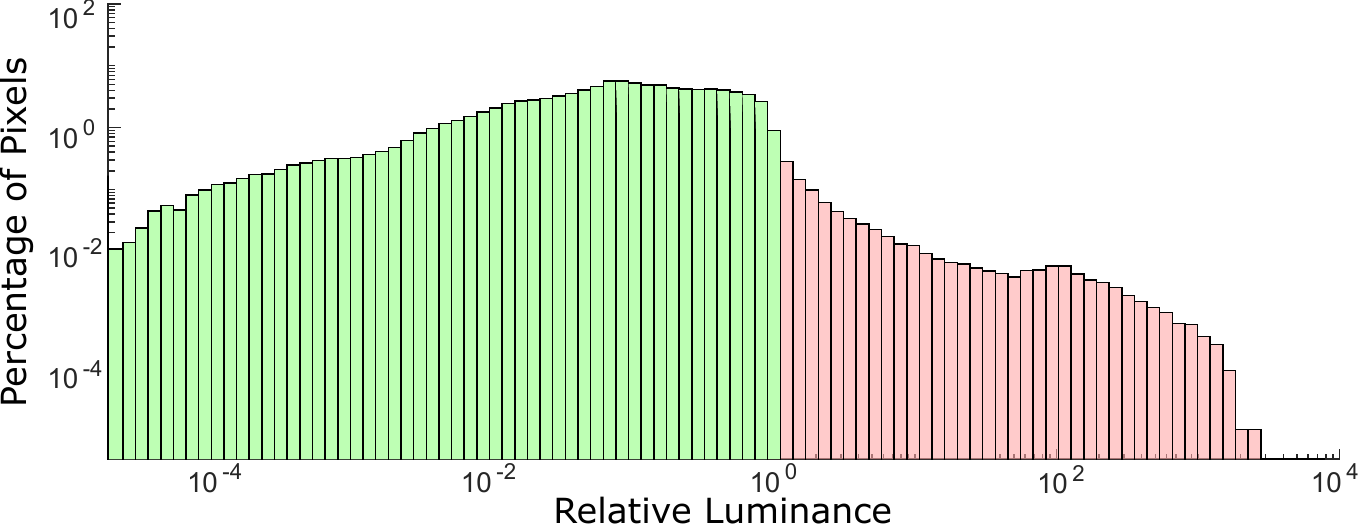}		
	\caption{Our training dataset consisted of HDR images scaled such that between 1 and 2\% of their pixels' values were saturated and clipped. Saturated pixels are shown in red.}
	\label{fig:DataAndHistogram}
\end{figure}

\begin{figure*}[t!]
	\centering
	\begin{subfigure}[b]{.32\textwidth}
		\centering
		{\includegraphics[width=.65\textwidth]{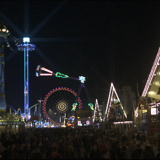}}
		~
		\begin{subfigure}[b]{.213\textwidth}	
			\cfbox{red}{\includegraphics[width=\linewidth]{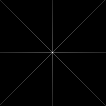}}
			\cfbox{green}{\includegraphics[width=\linewidth]{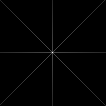}}
			\cfbox{blue}{\includegraphics[width=\linewidth]{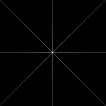}}
		\end{subfigure}
		\caption{Star PSF}
	\end{subfigure}%
	~
	\begin{subfigure}[b]{.32\textwidth}
		\centering
		\includegraphics[width=.65\textwidth]{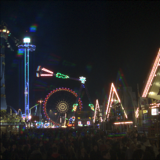}
		~
		\begin{subfigure}[b]{.213\textwidth}	
			\cfbox{red}{\includegraphics[width=\linewidth]{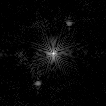}}
			\cfbox{green}{\includegraphics[width=\linewidth]{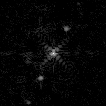}}
			\cfbox{blue}{\includegraphics[width=\linewidth]{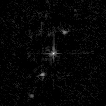}}
		\end{subfigure}
		\caption{E2E PSF}
	\end{subfigure}%
	~
	\begin{subfigure}[b]{.32\textwidth}
		\centering
		\includegraphics[width=.65\textwidth]{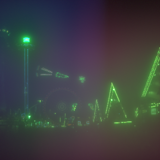}
		~
		\begin{subfigure}[b]{.213\textwidth}	
			\cfbox{red}{\includegraphics[width=\linewidth]{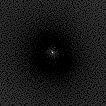}}
			\cfbox{green}{\includegraphics[width=\linewidth]{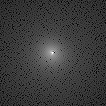}}
			\cfbox{blue}{\includegraphics[width=\linewidth]{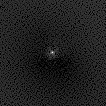}}
		\end{subfigure}
		\caption{E2E PSF (unconstrained)}
	\end{subfigure}%
	\centering
	\caption{Simulated sensor images for an example scene from our evaluation set and point spread functions (PSFs) for several different optical coding approaches: Rouf et al.'s star-shaped PSF~\shortcite{Rouf:2011} (a), our end-to-end deep optics approach applied with physically realizable constraints (b), and our end-to-end deep optics approach without physically realizable constraints applied (c). All three color channels of the PSFs are shown separately in the log domain. Whereas the star PSF continuously blurs the image along several radial streaks (a), the deep optics approach creates a PSF with several distinct peaks, which result in image content being copied at chromatically varying distances and at different intensity scales (b). The unconstrained PSF mostly blurs the green color channel while focusing the red and blue channels. 
	}
	\label{fig:PSFs}
\end{figure*}

\subsection{CNN-based Image Reconstruction}
\label{sec:cnn}

To recover $\mathbf{x}$ from $\mathbf{y}$ we use a convolutional neural network based on the well-known U-Net architecture~\cite{Unet}. Specifically, our U-Net uses skip connections and has 5 scales with 4 consecutive downsampling operations (maxpool) and 4 consecutive upsampling operations (transposed convolutions initialized with bilinear filter weights). At each scale of the U-Net, we include one additional convolutional layer; each convolutional layer is followed by a rectified linear unit (ReLU). BatchNorm layers are used after each upsampling layer and after the final convolutional layer. This architecture is inspired by Eilertsen's work~\shortcite{HDRCNN} but slightly leaner, which resulted in faster convergence of the lens' surface profile. As illustrated in Figure~\ref{fig:SystemDiagram}, each network layer has 64 feature maps. 

\subsection{End-to-end Training Details}
\label{sec:cnntraining}

We jointly optimize the PSF and the CNN via the end-to-end (E2E) framework illustrated in Figure \ref{fig:SystemDiagram}. In particular, using thousands of HDR training images, we simulate (1) passing an HDR image through our optical system, (2) capturing a noisy and saturated LDR image with a sensor, and (3) reconstructing the HDR images with the CNN. We compute the corresponding loss and use Tensorflow's autodifferentiation capabilities to back-propagate the error and update the parameters $\theta$ of the CNN and the height map $\doe$ of the PSF.

\paragraph{\bf Loss Function} 

Following Kalantari and Ramamoorthi~\shortcite{LearningHDR}, we originally experimented with minimizing the $\ell_2$-loss between the tone-mapped reconstruction and ground-truth HDR images. While successful in training the CNN, this loss function caused the network to focus its efforts on the most overexposed and challenging images in the training data. Using this approach, a typical Dirac delta-type was found as a locally optimal solution for the PSF.


To encourage the development of more powerful PSFs, we instead minimize the sum, over all the batches, of per-batch, $\gamma$-corrected, $\ell_2$-loss:
\begin{equation}
	\mathcal{L}_{\text{Data}} = \sum_{\text{B}\subset\text{Batches}}{\Big \|}(\mathbf{x}_{\text{B}}+\epsilon)^\gamma-(\hat{\mathbf{x}}_{\text{B}}+\epsilon)^\gamma{\Big \|}_2,
	\label{eq:lossgroup}
\end{equation}
where $\gamma=1/2$ and $\epsilon$ is a small constant we add to avoid the non-differentiability around $0$. That is, we minimized the root mean-squared-error (RMSE) over batches, as opposed to the typical mean-squared-error (MSE) loss.

In the context of regression, sums of $\ell_2$-norms over groups encourage group-sparse solutions~\cite{SparseGroupLasso}. In our context, sums of $\ell_2$-norms make the network more robust to outliers, i.e. it allows the network to fail for the most challenging reconstructions so long as the $\ell_2$-norm of the error is small for most batches.



\paragraph{\bf Incorporating Fabrication Constraints} 


To ensure that the optimized height map can be manufactured, we clip its values to the maximum range during training and add an additional smoothness term on $\doe$ to prevent the resulting surface profile to include many discontinuities. Specifically, we add the loss
\begin{equation}
	\mathcal{L}_{\text{Reg}} = \nu\|\mathbf{D}*\doe \|^2_2,
\end{equation}
to the overall loss function, where $\mathbf{D}$ is a Laplacian filter and $\nu=10^9$ is a weighting parameter.


\paragraph{\bf Datasets} 


Following Eilertsen et al.~\shortcite{HDRCNN}, we use training and validation datasets consisting of 2837 HDR images drawn from a combination of videos and images from a number of sources. We performed data augmentation by cropping, rescaling, and adjusting hue and saturation. The final training set contains just under 60,000 different HDR images with a resolution of $320\times 320$ pixels. Our test set consists of 223 HDR images, of which 83 are still images and the rest frames drawn from every 10th frame of four separate video sequences (which were not used for training). To generate LDR/HDR image pairs, we simulated capturing LDR frames where we set the exposure such that between 1 and 2\% of the pixels were saturated. A histogram of the pixel values in our training data is shown in Figure~\ref{fig:DataAndHistogram}. 

\paragraph{\bf Miscellaneous Training Details} 

We trained the end-to-end model using the Adam optimizer with a minibatch size of 8. We applied an exponential learning rate decay with an initial rate of $0.0001$. We trained the network for 100 epochs, which took about 3 days on a Pascal Titan~X graphics processing unit.

Source code, trained models, and datasets will be made available.

%% file: sections/analysis.tex

\begin{figure*}[t!]
	\centering
	\vspace{1 cm}
	\begin{subfigure}[t]{.15\textwidth}
		\centering
		\begin{overpic}[width=\textwidth]{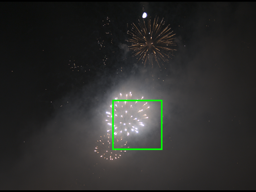}
		\put (20,90) {{\bf\centering\begin{tabular}{@{}c@{}} Reference \\ HDR \end{tabular}}}
		\end{overpic}
		\includegraphics[width=.48\textwidth]{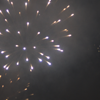}
	\end{subfigure}%
	~
	\begin{subfigure}[t]{.15\textwidth}
		\centering
		\captionsetup{justification=centering}
		\begin{overpic}[width=\textwidth]{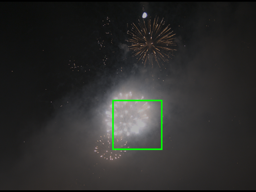}
			\put (35,80) {{\bf LDR}}
		\end{overpic}
		\includegraphics[width=.48\textwidth]{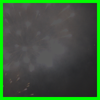}
		\includegraphics[width=.48\textwidth]{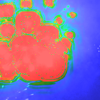}
		\caption*{PSNR = 46.2,\\Q = 43.5}
	\end{subfigure}
	~
	\begin{subfigure}[t]{.15\textwidth}
		\centering
		\captionsetup{justification=centering}
		\begin{overpic}[width=\textwidth]{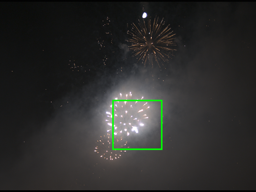}
			\put (15,80) {{\bf HDR-CNN}}
		\end{overpic}
		\includegraphics[width=.48\textwidth]{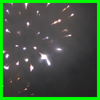}
		\includegraphics[width=.48\textwidth]{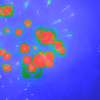}
		\caption*{PSNR = 47.9,\\Q = 49.1}
	\end{subfigure}
	~
	\begin{subfigure}[t]{.15\textwidth}
		\centering
		\captionsetup{justification=centering}
		\begin{overpic}[width=\textwidth]{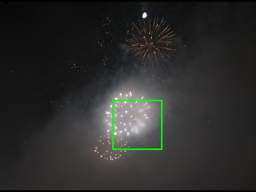}
			\put (34,80) {{\bf U-Net}}
		\end{overpic}
		\includegraphics[width=.48\textwidth]{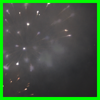}
		\includegraphics[width=.48\textwidth]{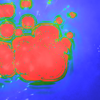}
		\caption*{PSNR = 47.7,\\Q= 43.7}
	\end{subfigure}
	~
	\begin{subfigure}[t]{.15\textwidth}
		\centering
		\captionsetup{justification=centering}
		\begin{overpic}[width=\textwidth]{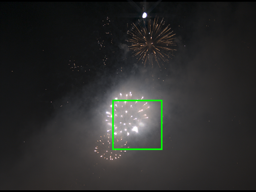}
			\put (10,90) {{\bf\centering\begin{tabular}{@{}c@{}} Star PSF + \\ U-Net \end{tabular}}}
		\end{overpic}
		\includegraphics[width=.48\textwidth]{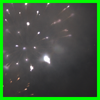}
		\includegraphics[width=.48\textwidth]{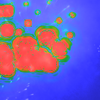}
		\caption*{PSNR = 48.0,\\Q = 44.8}
	\end{subfigure}
	~
	\begin{subfigure}[t]{.15\textwidth}
		\centering
		\captionsetup{justification=centering}
		\begin{overpic}[width=\textwidth]{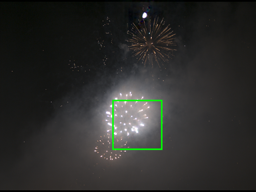}
			\put (10,90) {{\bf\centering\begin{tabular}{@{}c@{}} E2E PSF + \\ U-Net \end{tabular}}}
		\end{overpic}
		\includegraphics[width=.48\textwidth]{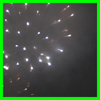}
		\includegraphics[width=.48\textwidth]{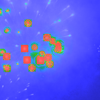}
		\caption*{PSNR = {\bf 50.7},\\Q = {\bf 53.7}}
	\end{subfigure}

	\begin{subfigure}[t]{.15\textwidth}
		\centering
		\captionsetup{justification=centering}
		\includegraphics[width=\textwidth]{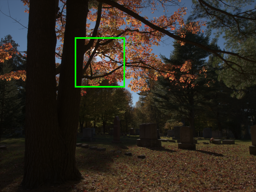}
		\includegraphics[width=.48\textwidth]{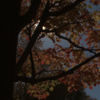}
	\end{subfigure}%
	~
	\begin{subfigure}[t]{.15\textwidth}
		\centering
		\captionsetup{justification=centering}
		\includegraphics[width=\textwidth]{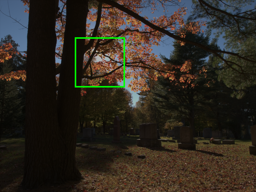}
		\includegraphics[width=.48\textwidth]{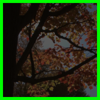}
		\includegraphics[width=.48\textwidth]{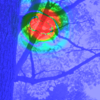}
		\caption*{PSNR = 51.5,\\Q = 51.7 }
	\end{subfigure}
	~
	\begin{subfigure}[t]{.15\textwidth}
		\centering
		\captionsetup{justification=centering}
		\includegraphics[width=\textwidth]{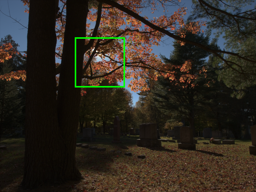}
		\includegraphics[width=.48\textwidth]{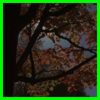}
		\includegraphics[width=.48\textwidth]{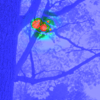}
		\caption*{PSNR = 53.4,\\Q = 57.5}
	\end{subfigure}
	~
	\begin{subfigure}[t]{.15\textwidth}
		\centering
		\captionsetup{justification=centering}
		\includegraphics[width=\textwidth]{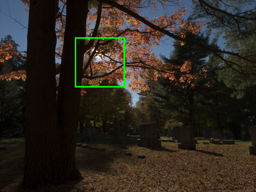}
		\includegraphics[width=.48\textwidth]{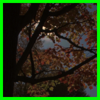}
		\includegraphics[width=.48\textwidth]{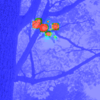}
		\caption*{PSNR = 55.0,\\Q = 61.1}
	\end{subfigure}
	~
	\begin{subfigure}[t]{.15\textwidth}
		\centering
		\captionsetup{justification=centering}
		\includegraphics[width=\textwidth]{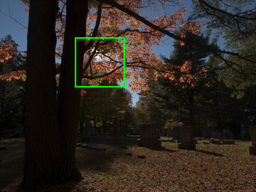}
		\includegraphics[width=.48\textwidth]{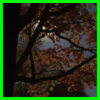}
		\includegraphics[width=.48\textwidth]{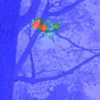}
		\caption*{PSNR = 55.2,\\Q = 62.3}
	\end{subfigure}
	~
	\begin{subfigure}[t]{.15\textwidth}
		\centering
		\captionsetup{justification=centering}
		\includegraphics[width=\textwidth]{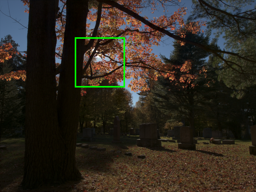}
		\includegraphics[width=.48\textwidth]{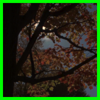}
		\includegraphics[width=.48\textwidth]{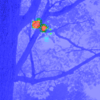}
		\caption*{PSNR = {\bf 57.7},\\Q = {\bf 65.7}}
	\end{subfigure}

	\begin{subfigure}[t]{.15\textwidth}
		\centering
		\captionsetup{justification=centering}
		\includegraphics[width=\textwidth]{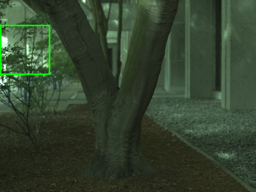}
		\includegraphics[width=.48\textwidth]{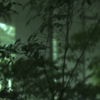}
	\end{subfigure}%
	~
	\begin{subfigure}[t]{.15\textwidth}
		\centering
		\captionsetup{justification=centering}
		\includegraphics[width=\textwidth]{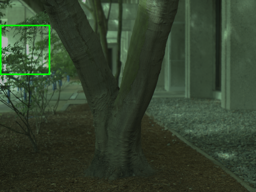}
		\includegraphics[width=.48\textwidth]{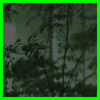}
		\includegraphics[width=.48\textwidth]{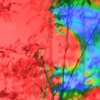}
		\caption*{PSNR = 41.3,\\Q = 46.9}
	\end{subfigure}
	~
	\begin{subfigure}[t]{.15\textwidth}
		\centering
		\captionsetup{justification=centering}
		\includegraphics[width=\textwidth]{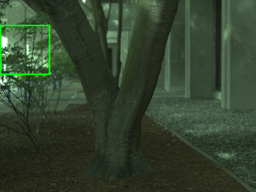}
		\includegraphics[width=.48\textwidth]{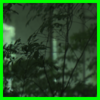}
		\includegraphics[width=.48\textwidth]{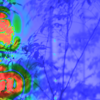}
		\caption*{PSNR = 45.9,\\Q = 57.1}
	\end{subfigure}
	~
	\begin{subfigure}[t]{.15\textwidth}
		\centering
		\captionsetup{justification=centering}
		\includegraphics[width=\textwidth]{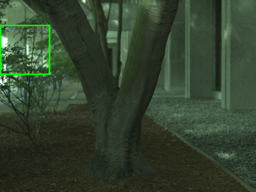}
		\includegraphics[width=.48\textwidth]{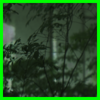}
		\includegraphics[width=.48\textwidth]{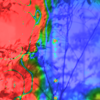}
		\caption*{PSNR = 44.1,\\Q = 53.3}
	\end{subfigure}
	~
	\begin{subfigure}[t]{.15\textwidth}
		\centering
		\captionsetup{justification=centering}
		\includegraphics[width=\textwidth]{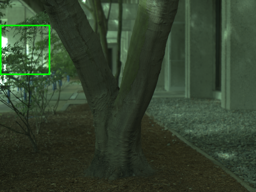}
		\includegraphics[width=.48\textwidth]{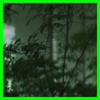}
		\includegraphics[width=.48\textwidth]{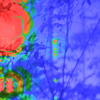}
		\caption*{PSNR = 44.4,\\Q = 56.2}
	\end{subfigure}
	~
	\begin{subfigure}[t]{.15\textwidth}
		\centering
		\captionsetup{justification=centering}
		\includegraphics[width=\textwidth]{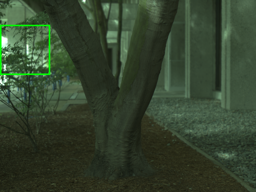}
		\includegraphics[width=.48\textwidth]{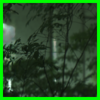}
		\includegraphics[width=.48\textwidth]{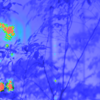}
		\caption*{PSNR = {\bf 48.5},\\Q = {\bf 59.9}}
	\end{subfigure}
\caption{Comparison of various single-shot HDR imaging approaches. In all examples, the whole images are displayed at -1 stop and regions of interest at -3 stops. In the columns, we show the ground truth HDR image, a corresponding LDR image, Eilertsen et al.'s CNN-based reconstruction applied to the LDR image (HDR-CNN)~\shortcite{HDRCNN}, a slightly simpler U-Net applied to the LDR image, the star-shaped PSF proposed by Rouf et al.~\shortcite{Rouf:2011} with a U-Net reconstruction, and our end-to-end deep optics approach (E2E). In all three examples, the peak signal-to-noise ratio (PSNR) and also observed image quality (see insets) are best for our approach. Color-coded insets show probabilities of perceiving the difference between the reconstructions and the ground truth HDR images, as computed by the HDR-VDP-2 visible differences predictor. Again, our approach qualitatively and quantitatively (evaluated with HDR-VDP-2 Q value) outperforms other approaches.}
\label{fig:SimComparo}
\end{figure*}



\rowcolors{2}{gray!25}{white}
\begin{table}[t!]%
	\begin{tabular}{l|cccc}
		\rowcolor{gray!50}
		\toprule
		& HDR-VDP-2 & PSNR-L & PSNR-$\gamma$ \\ 
		\midrule
		LDR     & 51.4 & 39.8 & 38.9 \\
		HDR-CNN \cite{HDRCNN} & 58.6 & 42.1 & 42.9  \\
		U-Net  & 56.4 & 41.8 & 42.3  \\
		\begin{tabular}{@{}l@{}}Star PSF + \\ U-Net\cite{Rouf:2011} \end{tabular}  & 56.7 & 42.1 & 42.3 \\	
		\begin{tabular}{@{}l@{}}E2E PSF +\\ U-Net \end{tabular} & 60.6 & 45.6 & {\bf 44.3} \\
		\begin{tabular}{@{}l@{}l@{}}E2E PSF +\\ U-Net\\(unconstrained)\end{tabular} & {\bf 67.1} & {\bf 46.8} & 40.7  \\
		\bottomrule
	\end{tabular}
	\caption{Quantitative evaluation for the entire test set. Several single-shot HDR imaging approaches are compared using a perceptual image difference computed by HDR-VDP-2 and peak signal-to-noise ratio (PSNR) computed in the linear domain (L) and in the $\gamma$-corrected domain.}
	\label{tab:SimResults}
\end{table}%

In this section, we evaluate the proposed method in simulation and show comparisons to various other single-shot HDR imaging approaches. 

Figure~\ref{fig:PSFs} shows simulated sensor images and PSFs for several options of optically coding the sensor image before reconstruction. The PSF that was optimized with the method proposed in the previous section is shown in (b) and also in Figure~\ref{fig:prototypelens}. This PSF contains several peaks, each creating a shifted and scaled copy of the sensor image superimposed on itself. In this way, the PSF serves to multiplex together different exposures of the image. The copies for individual color channels appear at slightly different location, which leads to visible chromatic aberrations in the sensor image; such chromatic aberrations are inevitable when using a single DOE design. We also optimize a PSF using the proposed optimization method but without parameterizing it by a physically realizable optical element (c). Therefore, the color channels of this PSF are completely independent of one another and individual pixel values can be arbitrary as long as the are in the range $[0,1]$ and sum to 1 for each channel. We also show the star-shaped PSF proposed by Rouf et al.~\shortcite{Rouf:2011}.

We compare several reconstruction approaches in Figure~\ref{fig:SimComparo}. These include the conventional LDR image, Eilertsen et al.'s CNN applied to this LDR image (HDR-CNN), the proposed smaller U-Net applied to this LDR image, the U-Net applied to an image captured with the star PSF, and our end-to-end deep optics approach with physically realizable constraints. For fair comparison, the U-Nets in each example are trained for the respective PSF. We show regions of interest in the insets along with visible differences predicted by HDR-VDP-2~\cite{Mantiuk:2011}. In all cases, the proposed deep optics approach achieves the best results.


Finally, we show a quantitative comparison of all these methods in Table~\ref{tab:SimResults}. Here, we report the average perceptual difference as computed by HDR-VDP-2 as well as peak signal-to-noise ratio (PSNR) over the entire test set in the linear and $\gamma$-corrected domains. We observed that the end-to-end (E2E) approaches achieve the best image quality. The unconstrained E2E approach is usually better than the physically realizable version because it has more degrees of freedom. However, the PSNR-$\gamma$ of the unconstrained PSF is slightly lower than that of the realizable approach. This is likely due to the fact that the network was trained using an outlier-robust loss, to which it over-fit. 
Recall, optimizing PSNR directly lead to sub-optimal convergence of the PSF (see Sec.~\ref{sec:cnntraining}).

%% file: sections/prototype.tex
\paragraph{\bf Lens Fabrication}

\begin{figure}[t!]
	\centering
		\includegraphics[width=\columnwidth]{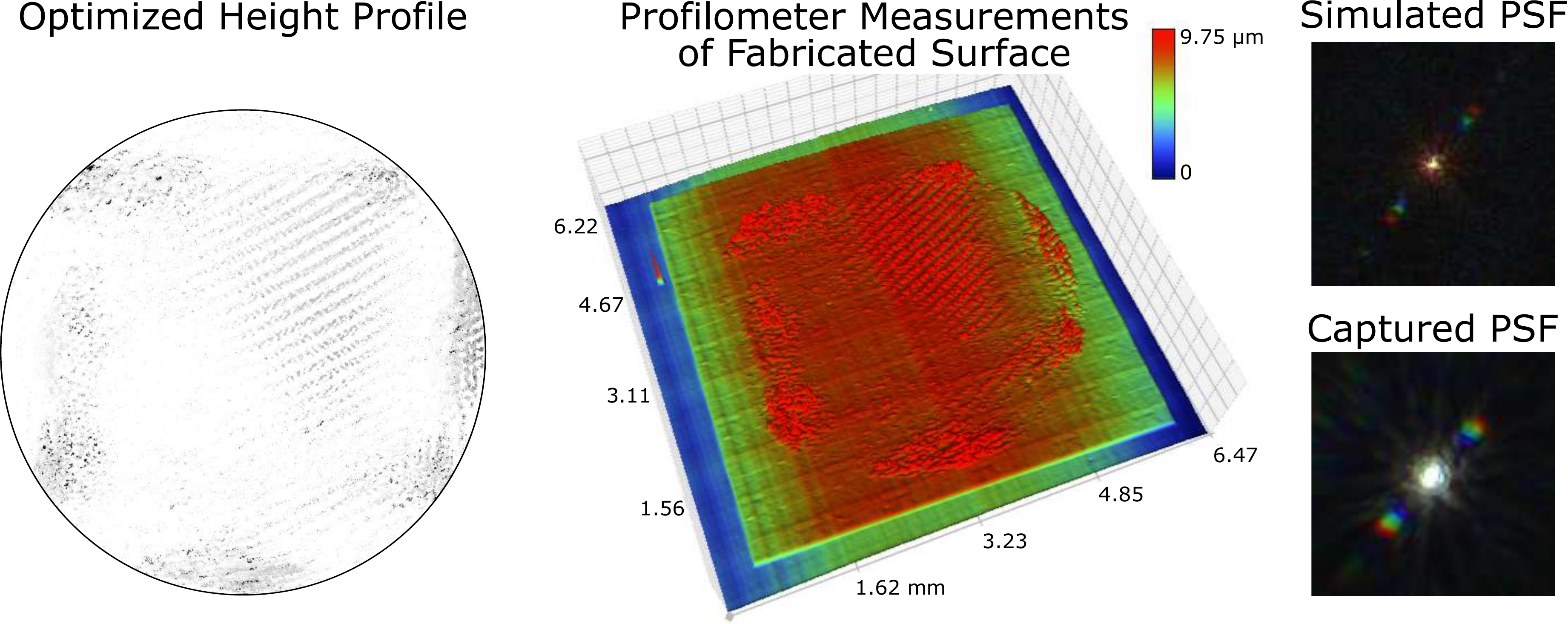}
		\caption{Optimized height profile of the diffractive optical element (DOE, left) along with profilometer measurements of the fabricated DOE. The DOE structure partially resembles that of a grating, which creates multiple peaks in the point spread function (PSF, right). Intuitively, this PSF creates three shifted and scaled copies of the input image. Although the measured PSF is slightly blurrier than the simulated PSF, likely due to imperfections in the fabrication process and approximations of our image formation model, their general shapes are comparable.}
		\label{fig:prototypelens}
\end{figure}

Once a phase profile is optimized, we fabricate the corresponding diffractive optical element (DOE) using polydimethyl-siloxane (PDMS) through replica molding. Figure~\ref{fig:prototypelens} shows the optimized height profile (left) along with a 3D rendering of profilometer measurements of the fabricated DOE (center), which has a diameter of 5~mm. Qualitatively, the shapes are similar. We also show the simulated (top right) and the captured PSFs (bottom right). These PSFs match well and both create a Dirac peak in the center and lower-amplitude satellite peaks at slightly different locations for the three color channels. This PSF is created by the lens surface profile that resembles a grating-like structure. The captured PSF is slightly blurrier than the simulation and there is additional glare, both likely due to slight fabrication errors and interreflections between the lens elements. 

\paragraph{\bf System Integration}

We mount the fabricated DOE as an add-on to a conventional single lens reflex camera (Canon Rebel T5) equipped with a standard compound lens (Nikon Nikkor 35~mm). The DOE is fixed in a Thorlabs lens tube with rotating optic adjustment (SM1M05), which is coupled to the SLR lens via an optomechanical adapter (Thorlabs SM1A2). Figure~\ref{fig:teaser} (right) shows the DOE and SLR lens. In this setup, the DOE is physically mounted at a slight distance to the compound camera lens. The exact distance between these optical elements is unknown because we model the primary camera lens as having a single refractive surface. While this is the easiest approach, a more detailed optical model of the compound lens may be desirable, although to model it appropriately, proprietary information from the lens manufacturer would have to be known. Moreover, the lack of anti-reflection coatings on the DOE may add interreflections and glare, and likely contributes to a slight mismatch between simulated and captured PSFs. 

To calibrate our camera system, we capture several different exposures of a white light source behind a \SI{75}{\micro\meter}-sized pinhole. These photographs are merged into a single HDR image using the technique described by Debevec~\shortcite{Debevec:1997}. This captured point spread function of our optical system is used to refine the pre-trained CNN by optimizing its parameters for the fixed captured PSF, as described in Section~\ref{sec:cnntraining}. Refining the CNN with a fixed PSF is significantly faster than training the end-to-end system from scratch and only takes a few hours.


%% file: sections/results.tex
\begin{figure*}[t!]
	\vspace{1 cm}
		\centering
	\begin{subfigure}[b]{.15\textwidth}
		\centering
		\begin{overpic}[width=\textwidth]{figures/ExperimentalRecons/Kitchen_teaser/GT_0_stop.png}
			\put (30,105) {{\bf LDR}}
		\end{overpic}
		\caption*{0 EV}
	\end{subfigure}%
	~
	\begin{subfigure}[b]{.15\textwidth}
		\centering
		\captionsetup{justification=centering}
		\begin{overpic}[width=\textwidth]{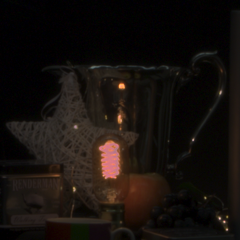}
			\put (10,105) {{\bf HDR-CNN}}
		\end{overpic}
		\caption*{-2 EV}
	\end{subfigure}
	~
	\begin{subfigure}[b]{.15\textwidth}
		\centering
		\captionsetup{justification=centering}
		\begin{overpic}[width=\textwidth]{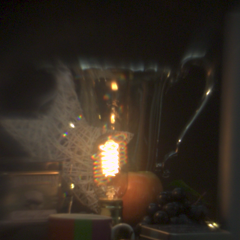}
			\put (5,113) {{\bf\centering\begin{tabular}{@{}c@{}} E2E \\ Measurement \end{tabular}}}
		\end{overpic}
		\caption*{0 EV}
	\end{subfigure}
	~
	\begin{subfigure}[b]{.15\textwidth}
		\centering
		\captionsetup{justification=centering}
		\begin{overpic}[width=\textwidth]{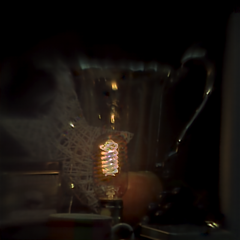}
			\put (0,113) {{\bf\centering\begin{tabular}{@{}c@{}} E2E \\ Reconstruction \end{tabular}}}
		\end{overpic}
		\caption*{-2 EV}
	\end{subfigure}
	~
	\begin{subfigure}[b]{.15\textwidth}
		\centering
		\captionsetup{justification=centering}
		\begin{overpic}[width=\textwidth]{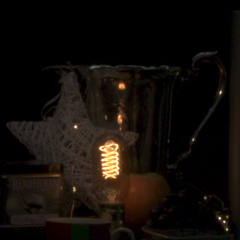}
			\put (-5,105) {{\bf Reference HDR}}
		\end{overpic}
		\caption*{-2 EV}
	\end{subfigure}
	~
	\begin{subfigure}[b]{.15\textwidth}
		\centering
		\captionsetup{justification=centering}
		\includegraphics[width=.48\textwidth]{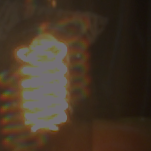}
		\includegraphics[width=.48\textwidth]{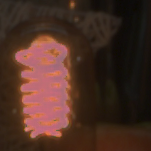}

		\includegraphics[width=.48\textwidth]{figures/ExperimentalRecons/Kitchen_teaser/reconMedNet_NewImbyImPSF_sig3_equalized_deconvolved_L2Gamma_HDR64_noise02_ROI_-232_stop.png}
		\includegraphics[width=.48\textwidth]{figures/ExperimentalRecons/Kitchen_teaser/GT_ROI_-232_stop.png}		
		\caption*{-2.3 EV}
	\end{subfigure}
	\begin{subfigure}[b]{.15\textwidth}
		\centering
		\begin{overpic}[width=\textwidth]{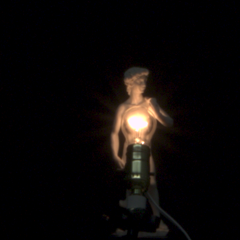}
		\end{overpic}
		\caption*{0 EV}
	\end{subfigure}%
	~
	\begin{subfigure}[b]{.15\textwidth}
		\centering
		\captionsetup{justification=centering}
		\begin{overpic}[width=\textwidth]{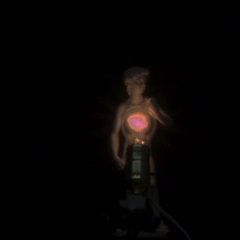}
		\end{overpic}
		\caption*{-2.3 EV}
	\end{subfigure}
	~
	\begin{subfigure}[b]{.15\textwidth}
		\centering
		\captionsetup{justification=centering}
		\begin{overpic}[width=\textwidth]{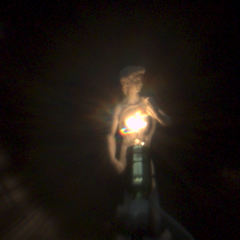}
		\end{overpic}
		\caption*{0 EV}
	\end{subfigure}
	~
	\begin{subfigure}[b]{.15\textwidth}
		\centering
		\captionsetup{justification=centering}
		\begin{overpic}[width=\textwidth]{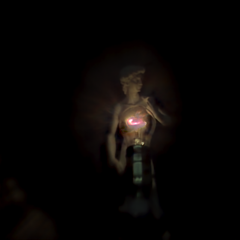}
		\end{overpic}
		\caption*{-2.3 EV}
	\end{subfigure}
	~
	\begin{subfigure}[b]{.15\textwidth}
		\centering
		\captionsetup{justification=centering}
		\begin{overpic}[width=\textwidth]{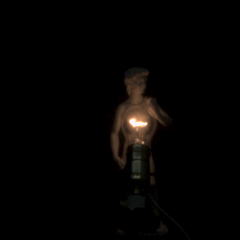}
		\end{overpic}
		\caption*{-2.3 EV}
	\end{subfigure}
	~
	\begin{subfigure}[b]{.15\textwidth}
		\centering
		\captionsetup{justification=centering}
		\includegraphics[width=.48\textwidth]{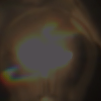}
		\includegraphics[width=.48\textwidth]{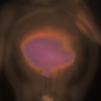}

		\includegraphics[width=.48\textwidth]{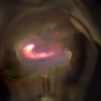}
		\includegraphics[width=.48\textwidth]{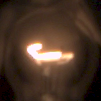}		
		\caption*{-3.3 EV}
	\end{subfigure}
	\begin{subfigure}[b]{.15\textwidth}
		\centering
		\begin{overpic}[width=\textwidth]{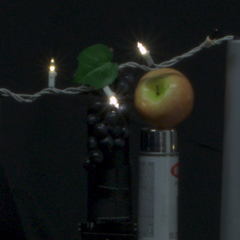}
		\end{overpic}
		\caption*{0 EV}
	\end{subfigure}%
	~
	\begin{subfigure}[b]{.15\textwidth}
		\centering
		\captionsetup{justification=centering}
		\begin{overpic}[width=\textwidth]{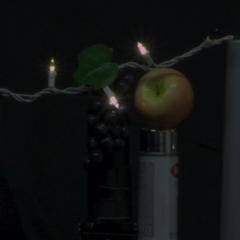}
		\end{overpic}
		\caption*{-1.3 EV}
	\end{subfigure}
	~
	\begin{subfigure}[b]{.15\textwidth}
		\centering
		\captionsetup{justification=centering}
		\begin{overpic}[width=\textwidth]{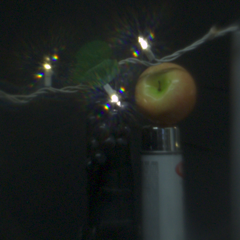}
		\end{overpic}
		\caption*{0 EV}
	\end{subfigure}
	~
	\begin{subfigure}[b]{.15\textwidth}
		\centering
		\captionsetup{justification=centering}
		\begin{overpic}[width=\textwidth]{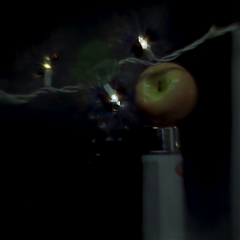}
		\end{overpic}
		\caption*{-1.3 EV}
	\end{subfigure}
	~
	\begin{subfigure}[b]{.15\textwidth}
		\centering
		\captionsetup{justification=centering}
		\begin{overpic}[width=\textwidth]{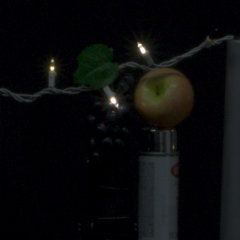}
		\end{overpic}
		\caption*{-1.3 EV}
	\end{subfigure}
	~
	\begin{subfigure}[b]{.15\textwidth}
		\centering
		\captionsetup{justification=centering}
		\includegraphics[width=.48\textwidth]{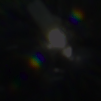}
		\includegraphics[width=.48\textwidth]{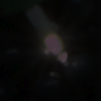}

		\includegraphics[width=.48\textwidth]{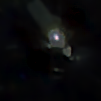}
		\includegraphics[width=.48\textwidth]{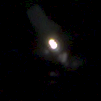}
		
		\caption*{-4.3 EV}
	\end{subfigure}
	\begin{subfigure}[b]{.15\textwidth}
	\centering
	\begin{overpic}[width=\textwidth]{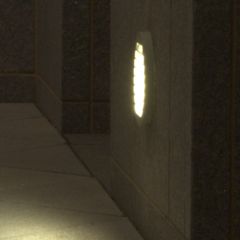}
	\end{overpic}
	\caption*{0 EV}
	\end{subfigure}%
	~
	\begin{subfigure}[b]{.15\textwidth}
		\centering
		\captionsetup{justification=centering}
		\begin{overpic}[width=\textwidth]{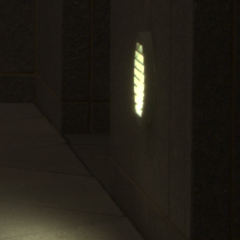}
		\end{overpic}
		\caption*{-1.6 EV}
	\end{subfigure}
	~
	\begin{subfigure}[b]{.15\textwidth}
		\centering
		\captionsetup{justification=centering}
		\begin{overpic}[width=\textwidth]{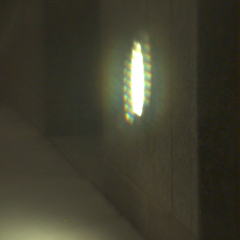}
		\end{overpic}
		\caption*{0 EV}
	\end{subfigure}
	~
	\begin{subfigure}[b]{.15\textwidth}
		\centering
		\captionsetup{justification=centering}
		\begin{overpic}[width=\textwidth]{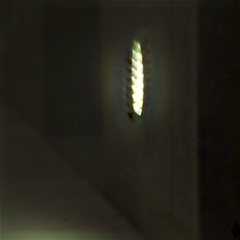}
		\end{overpic}
		\caption*{-1.6 EV}
	\end{subfigure}
	~
	\begin{subfigure}[b]{.15\textwidth}
		\centering
		\captionsetup{justification=centering}
		\begin{overpic}[width=\textwidth]{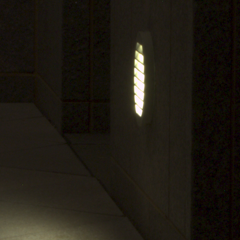}
		\end{overpic}
		\caption*{-1.6 EV}
	\end{subfigure}
	~
	\begin{subfigure}[b]{.15\textwidth}
		\centering
		\captionsetup{justification=centering}
		\includegraphics[width=.48\textwidth]{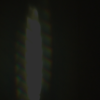}
		\includegraphics[width=.48\textwidth]{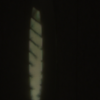}
		
		\includegraphics[width=.48\textwidth]{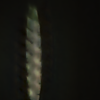}
		\includegraphics[width=.48\textwidth]{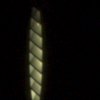}
		
		\caption*{-4.9 EV}
	\end{subfigure}
	\caption{Experimental results of three indoor scenes (top three rows) and one outdoor scene at night (bottom row). The limited dynamic range of the sensor loses details in the brighter parts of the captured LDR image (first column) as compared with the reference HDR image (fifth column). A CNN operating directly on the LDR images hallucinates brighter content in these saturated parts, but it is missing the detail (second column). The measurements captured with our prototype camera optically encode this detail in the image by superimposing several scaled and shifted copies of the image on itself (third column). This information is used by our CNN to recover the missing parts of the scene while digitally removing the image copies (fourth column). The closeups, showing E2E measurements, HDR-CNN, E2E Reconstruction, and ground truth HDR, demonstrate that deep optics is more successful in recovering bright detail of HDR scenes than other single-shot HDR imaging approaches (right column).}
	\label{fig:ExpComparo}
\end{figure*}

Using the prototype camera described in the previous section, we captured several HDR example scenes (see Figs.~\ref{fig:teaser},~\ref{fig:ExpComparo}). These include three scenes recorded in a laboratory setting (Fig.~\ref{fig:teaser} and Fig.~\ref{fig:ExpComparo}, top and center row) and one outdoor scene captured at night (Fig.~\ref{fig:ExpComparo}, bottom row). In Figure~\ref{fig:ExpComparo}, captured measurements along with reconstructions computed by our CNN as well as reference LDR and HDR images and the result of the HDR-CNN~\cite{HDRCNN} are shown. In all of these examples, the captured LDR images include saturated areas that actually contain detail, which is completely lost in the measurements, such as the filament of a light bulb (top row) or the structure of a light source on a wall (bottom row). We show these images as well as magnified closeups (right column) at varying exposure values (EVs) to best highlight these details. 
In should be noted that these are all examples where we expect the HDR-CNN approach to fail, because the network simply has no information about the detail in the saturated parts---the best it can do is to inpaint these parts. The inpainting process results in smooth regions that exceed the dynamic range of the LDR sensor image but that do not resemble the actual content in these examples.


\rowcolors{1}{white}{white}

%% file: sections/discussion.tex

In summary, we propose an end-to-end approach to jointly training an optical encoder, i.e. the point spread function created by a custom optical element, and electronic decoder, i.e. a convolutional neural network, for the application of single-shot high-dynamic-range imaging. 
As opposed to CNN-based methods that operate directly on conventional LDR images, our deep optics approach has the ability to optically encode details from bright parts of the scene into the LDR measurements. 
In particular, our method uses a unique multiplexing solution to HDR imaging, which it developed automatically, wherein the PSF superimposes multiple shifted exposures on top of one another. 
The proposed framework builds on the emerging idea of end-to-end optimization of optics and image processing, but to our knowledge it is the first to explore this general methodology for single-shot HDR imaging.




\paragraph{\bf Limitations}

Conceptually, HDR-CNN approaches that operate directly on conventional LDR images solve an inpainting problem, which is ill-posed and often fails. Accordingly, and unlike our deep optics approach, the worst case solution of HDR-CNNs is a conventional LDR image, which is directly recorded and which may not always need additional post-processing to begin with. Our method changes the optical image formation, so post-processing becomes a necessary part of the imaging pipeline. Slight reconstruction artifacts in our experimental results (Figs.~\ref{fig:teaser},~\ref{fig:ExpComparo}), which are primarily due to imperfections in the PSF calibration, can be found in both non-saturated and saturated parts of the image due to the need for processing the entire LDR image, rather than just its saturated parts.

In essence, the inverse problem in our method is more closely related to deconvolution problems than to inpainting problems, although for extremely bright scenes where even the lower intensity copies of the sensor image saturate, inpainting is unavoidable and the additional scene copies in our measurements may actually be harmful. Therefore, careful curation of the training set is crucial, because it needs to include HDR images with values adequately representing those observed during inference. A network is likely going to fail to produce high-quality results for conditions that it has not been trained for.


The fabricated diffractive optical element creates a PSF that closely resembles the simulated PSF. Yet, blur and glare, likely due to interreflections between optical elements, are problematic. Optical blur makes the deconvolution problem harder and glare causes the PSF to be slightly shift variant, which limits the effective field of view of our captured data (depth-of-field is unaffected). Thus, although the add-on approach of mounting the DOE in front of an SLR camera lens is convenient and flexible, integrating the DOE into the aperture plane of the primary lens may produce better results, as it more closely resembles our paraxial image formation model. Alternatively, the image formation model could be generalized to a non-paraxial model. Finally, anti-reflection coatings may help mitigate glare in the optics. 

\paragraph{\bf Future Work}

The end-to-end methods enable designing tailored optics for a particular task, rather than just capturing the sharpest image. Evaluating the benefits of end-to-end optimization of optics and image processing for other applications, including multispectral imaging, light field imaging, lensless imaging, and computational microscopy, is an interesting avenue of future work. 

%% file: sections/conclusion.tex
End-to-end design of optics and image reconstruction algorithms is an emerging paradigm for designing domain-specific computational cameras. We demonstrate clear benefits of this approach to the long-standing problem of single-shot HDR imaging. Beyond this application, deep computational cameras have the potential to re-define the next generation of domain-specific cameras that capture richer visual information than conventional cameras and that enable entirely new imaging modalities in a wide range of applications.